\documentclass[
 sort&compress
    ,final            % use final for the camera ready runs
 ,numberedheadings % uncomment this option for numbered sections
  ]
  {aipproc}

\layoutstyle{6x9}

\usepackage[latin1]{inputenc}
\usepackage[T1]{fontenc}
\usepackage{amsmath}
\usepackage{hyperref}

\newcommand{\eref}[1]{eq.~(\ref{#1})}

\newcommand{\nnnl}{\nonumber\\}	%nonumber new line

\newcommand{\vp}[0]{\varphi} % phi
\newcommand{\gam}[0]{\hat{\gamma}}

\begin{document}

\title{Gribov horizon and BRST symmetry: a pathway to confinement}


\author{S. P. Sorella\footnote{Speaker at the XII Mexican Workshop on Particles and
Fields, 9-14 November 2009, Mazatlan, Sinaloa, Mexico.}\,\,}{
  address={UERJ - Universidade do Estado do Rio de Janeiro, Instituto de F\'isica - Departamento de F\'isica Te\'orica, Rua S\~ao Francisco Xavier 524, 20550-013 Maracan\~a,
Rio de Janeiro, Brasil}
}

\author{L. Baulieu}{
  address={Theory Division, CERN, 1211-Geneve 23, Switzerland},
  altaddress={LPTHE, Universit\'{e}s Pierre et Marie Curie, 4 place Jussieu, F-75252 Paris Cedex 05, France}
}

\author{D. Dudal}{
  address={Ghent University, Department of Physics and Astronomy, Krijgslaan 281-S9, B-9000 Gent, Belgium}
}

\author{M. S. Guimaraes}{
  address={UERJ - Universidade do Estado do Rio de Janeiro, Instituto de F\'isica - Departamento de F\'isica Te\'orica, Rua S\~ao Francisco Xavier 524, 20550-013 Maracan\~a,
Rio de Janeiro, Brasil}
}

\author{M. Q. Huber}{
  address={Institut f\"ur Physik, Karl-Franzens-Universit\"at Graz, Universit\"atsplatz 5, 8010 Graz, Austria}
}

\author{N. Vandersickel}{
  address={Ghent University, Department of Physics and Astronomy, Krijgslaan 281-S9, B-9000 Gent, Belgium}
}

\author{D. Zwanziger}{
  address={New York University, New York, NY 10003, USA}
}

\begin{abstract}
 We summarize the construction of the Gribov-Zwanziger action and how it leads to a scenario which explains the confinement of gluons, in the sense that the elementary gluon excitations violate positivity. Then we address the question of how one can construct operators within this picture whose one-loop correlation functions have the correct analytic properties in order to correspond to physical excitations. For this we introduce the concept of $i$-particles.
\end{abstract}

\maketitle

The absence of quarks and gluons from the physical spectrum is known as confinement. To understand how it works is quite a challenge and has been investigated for decades. As a first step to better comprehend the full theory of the strong interaction, quantum chromodynamics, the case of Yang-Mills theory may be analyzed.

The action of Yang-Mills theory in Euclidean space is given by
\begin{align}\label{eq:S_YM}
S_{YM}&=\frac1{4}\int d^4 x F_{\mu\nu}^{a}F_{\mu\nu}^{a},
\end{align}
where $F_{\mu\nu}^{a}$ is the field strength tensor. This action is invariant under local SU($N$) gauge transformations. However, in the path integral one aims at integrating only over physically inequivalent configurations. To achieve this one fixes the gauge. In the following we will choose
the Landau gauge, which amounts to constrain the fields to satisfy $\partial_\mu A_\mu=0$. Properly implementing this restriction leads to the following additional term in the action:
\begin{align}\label{eq:S_gf}
 S_{gf}&=\int d^4 x\;  (i\,b^a \partial_\mu A_\mu^a-\bar{c}^a {\cal M}^{ab} c^b).
\end{align}
$b$ is the Lagrange multiplier field to enforce the gauge fixing condition and $c$ and $\bar{c}$ are the Faddeev-Popov ghosts, which are introduced in order to obtain a local action from the gauge fixing procedure. ${\cal M}^{ab}=-\partial_\mu D_\mu^{ab}  = -(\delta^{ab}\partial^2+g\,f^{abc} A_\mu^c \partial_\mu)$ denotes the Faddeev-Popov operator. The new action $S_{YM}+S_{gf}$ can be used as long as fluctuations around the origin, $A=0$, are small, i.~e. in perturbation theory. However, in the non-perturbative regime new physically equivalent configurations are encountered, the so-called Gribov copies \cite{Gribov:1977wm}. This is an intrinsic problem of gauge theories and cannot be circumvented for example by choosing another gauge \cite{Singer:1978dk}. In order to remedy this problem Gribov suggested a further modification of the gauge fixing, namely to restrict the domain of integration in field space to what is nowadays called the first Gribov region $\Omega$ \cite{Gribov:1977wm}. It is defined as the set of Landau gauge configurations for which the Faddeev-Popov operator is positive, i.~e.
\begin{align}
\Omega = \{ \; A; \; \partial_{\mu} A^a_{\mu}=0, \; {\cal M}^{ab} > 0 \; \}.  \label{om}
\end{align}
This region is convex, bounded in all directions, contains the origin in field space,  and all gauge orbits pass through it \cite{Dell'Antonio:1991xt,Zwanziger:2003cf}. Its boundary, called the Gribov horizon, is defined by the vanishing of the lowest eigenvalue of the Faddeev-Popov operator. A restriction to this region can be implemented via the addition of a non-local term to the action \cite{Zwanziger:1989mf, Zwanziger:1993dh}, the so-called horizon term:
\begin{align}
S_h =\int d^4 x \,h(x) =   \int d^4 x \lim_{\gamma(x)\to \gamma} \int  d^4 y  D_\mu^{ac}(x) \gamma^2(x) (\mathcal M^{-1})^{ad}(x,y) D_\mu^{dc} \gamma^2(y).   \label{ho}
\end{align}
Here a new parameter $\gamma$ with dimension of mass occurs. It is not free, but determined by the horizon condition $ \langle h(x) \rangle =d \gamma^4 (N^2-1)$ \cite{Zwanziger:1989mf},
where $N$ is the number of colors and $d$ the space-time dimension. The horizon term can be localized by the introduction of new fields \cite{Zwanziger:1989mf}: $(\bar{\vp}_\mu^{ab}, \vp_\mu^{ab}, \omega_\mu^{ab}, \bar{\omega}_\mu^{ab})$. The former two are a pair of complex bosonic fields and the latter two a pair of complex fermionic fields. This leads to the Gribov-Zwanziger action
\begin{align}\label{eq:S_GZ}
S_{GZ}&=S_{YM}+S_{gf}+S_{loc}\\
S_{loc}&=\int d^4x\Big( \bar{\vp}_\mu^{ac} {\cal M}^{ab} \vp_\mu^{bc}-\bar{\omega}_\mu^{ac} {\cal M}^{ab} \omega_\mu^{bc}
-g\,f^{abc}(\partial_\nu\bar{\omega}_\mu^{ad})(D^{be}_\nu c^e)\vp_\mu^{cd}+\nnnl
 &\qquad +\gamma^2\,g\,f^{abc}A_\mu^{abc}(\vp_\mu^{bc}-\bar{\vp}_\mu^{bc})-d(N^2-1)\gamma^4 \Big) . \label{locact12}
\end{align}
The last term is introduced in order to be able to rewrite the horizon condition as
$\frac{\delta\Gamma}{\delta \gamma^2}=0$,%  \label{geq}
where $\Gamma$ is the vacuum energy.

An important aspect of the Gribov-Zwanziger action is that confinement is already manifest at the perturbative level. For this consider the tree-level propagator of the gluon:
\begin{align}
\left\langle A_{\mu }^{a}(k)A_{\nu }^{b}(-k)\right\rangle =\delta
^{ab}\left( \delta _{\mu \nu }-\frac{k_{\mu }k_{\nu }}{k^{2}}\right) \frac{k^{2}}{k^{4}+{\hat \gamma }^{4}}, \qquad {\hat{\gamma}}^4 = 2 g^2 N \gamma^4. \label{z11}
\end{align}
Its poles are at $k^{2}=\pm i{\hat \gamma}^{2}$, so they do not correspond to physical excitations. This is also evident as the propagator \eref{z11} has negative norm contributions and hence violates positivity \cite{Alkofer:2003jj}. Thus in this scenario gluons are confined by the presence of the Gribov horizon. The ghost propagator is infrared enhanced due to the horizon condition and goes like $1/k^4$ \cite{Zwanziger:1993dh,Zwanziger:1992qr}.
The same qualitative results emerge from the non-perturbative Dyson-Schwinger and functional renormalization group equations \cite{vonSmekal:1997vx,Pawlowski:2003hq}. Even taking explicitly into account the Gribov horizon does not change this \cite{Zwanziger:2001kw,Huber:2009tx}.
Furthermore, at first sight this is in agreement with the Kugo-Ojima confinement scenario \cite{Kugo:1995km,Kugo:1979gm}. Taking the Kugo-Ojima confinement condition as constraint for the Yang-Mills action indeed leads to a similar term as the horizon term \cite{Dudal:2009xh}. It must be pointed out though that a different picture seems to emerge from recent lattice calculations \cite{Cucchieri:2007md}: the positivity violating gluon propagator is infrared suppressed and non-vanishing at zero momentum, while the ghost propagator is no longer enhanced in the infrared, behaving essentially as $\frac{1}{k^2}$ for $k \approx 0$. This behavior has been shown to be accommodated for in the Gribov-Zwanziger action by taking into account dynamical effects related to the condensation of local dimension two operators. This leads to the so called Refined Gribov-Zwanziger model \cite{Dudal:2008sp}.

Gauge symmetry, once fixed, is replaced by another, very useful symmetry called BRST symmetry. It can be used to prove the renormalizability of the action and to define the physical subspace of the theory. Nevertheless, restricting the domain of integration in field space even further, as realized by the Gribov-Zwanziger action, \eref{eq:S_GZ}, leads to a BRST symmetry which is softly  broken. Since the  breaking is soft, the action remains multiplicative renormalizable \cite{Zwanziger:1992qr,Maggiore:1993wq,Dudal:2005na}, but the issue of how to define the physical subspace remains to be clarified. However, it is possible to rewrite the broken symmetry into an exact non-local symmetry of the Gribov-Zwanziger action \cite{Sorella:2009vt}. Also the meaning of the Kugo-Ojima criterion becomes obscure by the lack of BRST invariance \cite{Dudal:2008sp,Dudal:2009xh}.

\paragraph{Finding physical operators}
The physical observable quantities of Yang-Mills theory are not gluons but glueballs. They should be described by gauge invariant operators whose correlation functions exhibit good analyticity properties in the complex cut Euclidean $k^2$-plane: i.e.  poles and cuts located on the negative real axis as well as a spectral representation with positive spectral density. The simplest candidate is the operator $O(x)= F_{\mu\nu}^2(x)$. In \cite{Zwanziger:1989mf} the analytic properties of its correlation function were investigated at one-loop order, using the Gribov-Zwanziger action. The promising outcome was that the correlation function contains physical and unphysical parts. In the present context we use the word physical for an operator with a cut on the negative real axis and a positive spectral function. The unphysical case corresponds to cuts starting somewhere else in the complex plane.

In order to get rid of the unphysical part one can try to deform the operator $O(x)$ appropriately. For this it turns out to be useful to diagonalize the conventional Gribov-Zwanziger action, \eref{locact12}, and construct operators from the resulting fields \cite{Baulieu:2009ha}. We will restrict ourselves to the quadratic part and perform all calculations at leading order in perturbation theory. For the diagonalization of the Lagrangian we first split the fields $\vp$ and $\bar{\vp}$ into real and imaginary parts:
\begin{align}
\vp^{ab}_{\mu} = \frac{1}{\sqrt{2}} \left( U^{ab}_{\mu} + i\, V^{ab}_{\mu} \right),
\qquad  {\bar \vp}^{ab}_{\mu} = \frac{1}{\sqrt{2}} \left( U^{ab}_{\mu} - i\, V^{ab}_{\mu} \right). \label{vu}
\end{align}
In the resulting action we observe that the gluon field mixes only with the adjoint part of $V_\mu^{ab}$, given by
$V^{p}_{\mu}= \frac{1}{N} f^{pmn} V^{mn}_\mu$. %\label{adjp}
Hence we decompose the $V$-field as
\begin{align}
V^{ab}_{\mu}= \frac{1}{N} f^{abp}f^{pmn} V^{mn}_{\mu} + \left( V^{ab}_{\mu} - \frac{1}{N} f^{abp}f^{pmn} V^{mn}_{\mu} \right) = f^{abp} V^{p}_{\mu} + S^{ab}_{\mu}. \label{dec}
\end{align}
A complete diagonalization is achieved by introducing the fields $\lambda_\mu^a$ and $\eta_\mu^a$:
\begin{align}
A^{a}_{\mu}  = \frac{1}{\sqrt{2}} \left( \lambda^{a}_{\mu}+ \eta^{a}_{\mu} \right),
\qquad {V}^{a}_{\mu}  = \frac{1}{\sqrt{2N}} \left( \lambda^{a}_{\mu}- \eta^{a}_{\mu} \right). \label{fv}
\end{align}
Finally the quadratic part of the action is then
\begin{align}
S_{GZ}^{\rm quad} = \int d^4x & \left( \frac{1}{2} {\lambda}^{a}_{\mu} \left(-\partial^2+ i\sqrt{2N}g\gamma^2 \right) {\lambda}^{a}_{\mu} + \frac{1}{2} {\eta}^{a}_{\mu}\left(-\partial^2- i\sqrt{2N}g\gamma^2\right){\eta}^{a}_{\mu}  \right)\nonumber \\
+ \int d^4x & \left( \frac{1}{2} S^{ab}_{\mu} (-\partial^2) S^{ab}_{\mu} +  \frac{1}{2} U^{ab}_{\mu}(-\partial^2)U^{ab}_{\mu} -\bar{\omega}_\mu^{ac} {\cal M}^{ab} \omega_\mu^{bc}\right), \label{quad4}
\end{align}
where the Landau condition $\partial_\mu A_\mu=0$ was used. The propagators of the new fields are
\begin{align}
\langle \lambda^{a}_{\mu}(k) \lambda^{b}_{\nu}(-k) \rangle & =  \frac{\delta^{ab}}{k^2  + i \hat{\gamma}^2} \left(\delta_{\mu\nu} -\frac{k_{\mu}k_{\nu}}{k^2} \right) , \nonumber \\
\langle \eta^{a}_{\mu}(k) \eta^{b}_{\nu}(-k) \rangle & = \frac{\delta^{ab}}{k^2  - i \hat{\gamma}^2} \left(\delta_{\mu\nu} -\frac{k_{\mu}k_{\nu}}{k^2} \right). \label{letp}
\end{align}
$\eta_\mu^a$ and $\lambda_\mu^a$ are called $i$-particles, as their poles are at the unphysical values $\pm  i \hat{\gamma}^2$. One should remember that the gluon propagator \eref{z11} could be written as
\begin{align}
\delta
^{ab}\left( \delta _{\mu \nu }-\frac{k_{\mu }k_{\nu }}{k^{2}}\right) \frac{k^{2}}{k^{4}+{\hat \gamma }^{4}}=\delta
^{ab}\left( \delta _{\mu \nu }-\frac{k_{\mu }k_{\nu }}{k^{2}}\right) \frac{1}{2} \left( \frac{1}{k^2-i{\hat \gamma}^2} + \frac{1}{k^2+i {\hat \gamma}^2 }\right).
\end{align}
This corresponds to the propagation of two unphysical particles with poles at $\pm  i \hat{\gamma}^2$ which can be identified with the $i$-fields.

With these new fields one can easily construct a physical operator. Introducing
\begin{align}
\lambda^{a}_{\mu\nu} = \partial_{\mu}  \lambda^{a}_{\nu} -  \partial_{\nu}  \lambda^{a}_{\mu},
\qquad \eta^{a}_{\mu\nu} = \partial_{\mu}  \eta^{a}_{\nu} -  \partial_{\nu}  \eta^{a}_{\mu} \label{fs}
\end{align}
as the $i$-field strengths, two simple examples are
\begin{align}
O^{(1)}_{\lambda\eta}(x) =  \left( \lambda^{a}_{\mu\nu}(x) \eta^{a}_{\mu\nu}(x) \right),
\qquad O^{(2)}_{\lambda\eta}(x) =  \varepsilon_{\mu\nu\rho\sigma} \left( \lambda^{a}_{\mu\nu}(x) \eta^{a}_{\rho\sigma(x)} \right).
\end{align}
The correlation function of the first in four dimensions is \cite{Baulieu:2009ha}
\begin{align}
\label{GZres2}  \langle& O^{(1)}_{\lambda\eta}(k) O^{(1)}_{\lambda\eta}(-k) \rangle = 12 (N^2-1) \int_{2\gam^2}^\infty d\tau \frac{1}{\tau+k^2} \frac{ \sqrt{\tau^2-4 \gam^4}(2\gam^4+\tau^2)}{32   \pi ^2 \tau}.
\end{align}
The employed K\"all\'{e}n-Lehmann representation nicely exhibits the cut from $-2\hat{\gamma}^2$ to $-\infty$ and the spectral density is positive.

\paragraph{Conclusions}

The Gribov-Zwanziger action is obtained by an improved gauge fixing. The resulting tree-level gluon propagator describes confined gluons and the ghost propagator is infrared enhanced. For the construction of operators corresponding to physical excitations of the theory we introduced the useful concept of $i$-particles. However, although the operators analyzed possess only cuts along the negative real axis and positive spectral functions, they constitute just a first small step towards a description of glueballs as some challenges have yet to be faced like the non-trivial extension of these operators to the quantum level, a necessary step in order to perform higher loops calculations.  Due to the soft breaking of the BRST symmetry, this point requires special care as one learns from the renormalization of the operator $F_{\mu\nu}^aF_{\mu\nu}^a$ \cite{Dudal:2009zh}, where both  BRST exact and BRST non-invariant quantities are needed in order to construct a quantum operator invariant under the renormalization group equations.

\begin{theacknowledgments}
D.~Dudal and N.~Vandersickel are supported by the Research-Foundation
Flanders (FWO Vlaanderen). The Conselho Nacional de Desenvolvimento Cient\'{\i}fico e Tecnol\'{o}gico (CNPq-Brazil),
the Faperj, Funda{\c{c}}{\~{a}}o de Amparo {\`{a}} Pesquisa do Estado do Rio
de Janeiro, the SR2-UERJ and the Coordena{\c{c}}{\~{a}}o de Aperfei{\c{c}}%
oamento de Pessoal de N{\'{\i}}vel Superior (CAPES) are gratefully
acknowledged for financial support. M.~Q.~Huber is supported by the FWF under contract W1203-N08.
\end{theacknowledgments}


\bibliographystyle{aipproc}   % if natbib is available

\begin{thebibliography}{99}

\bibitem[Gribov(1978)]{Gribov:1977wm}
V.~N. Gribov, \emph{Nucl. Phys.} \textbf{B139}, 1 (1978).

\bibitem[Singer(1978)]{Singer:1978dk}
I.~M. Singer, \emph{Commun. Math. Phys.} \textbf{60}, 7 (1978).

\bibitem[Dell'Antonio and Zwanziger(1991)]{Dell'Antonio:1991xt}
G.~Dell'Antonio, and D.~Zwanziger, \emph{Commun. Math. Phys.} \textbf{138}, 291 (1991).

\bibitem[Zwanziger(2004)]{Zwanziger:2003cf}
D.~Zwanziger, \emph{Phys. Rev.} \textbf{D69}, 016002 (2004).

\bibitem[Zwanziger(1989)]{Zwanziger:1989mf}
D.~Zwanziger, \emph{Nucl. Phys.} \textbf{B323}, 513 (1989).

\bibitem[Zwanziger(1994)]{Zwanziger:1993dh}
D.~Zwanziger, \emph{Nucl. Phys.} \textbf{B412}, 657 (1994).

\bibitem[Alkofer et~al.(2004)]{Alkofer:2003jj}
R.~Alkofer, W.~Detmold, C.~S. Fischer, and P.~Maris, \emph{Phys. Rev.} \textbf{D70}, 014014 (2004).

\bibitem[Zwanziger(1993)]{Zwanziger:1992qr}
D.~Zwanziger, \emph{Nucl. Phys.} \textbf{B399}, 477 (1993).

\bibitem[von Smekal et~al.(1998)]{vonSmekal:1997vx}
L.~von Smekal, A.~Hauck, and R.~Alkofer, \emph{Ann. Phys.} \textbf{267}, 1  (1998).

\bibitem[Pawlowski et~al.(2004)]{Pawlowski:2003hq}
J.~M. Pawlowski, D.~F. Litim, S.~Nedelko, and L.~von Smekal, \emph{Phys. Rev. Lett.} \textbf{93}, 152002 (2004).

\bibitem[Zwanziger(2002)]{Zwanziger:2001kw}
D.~Zwanziger, \emph{Phys. Rev.} \textbf{D65}, 094039 (2002).

\bibitem[Huber et~al.(2009)]{Huber:2009tx}
M.~Q. Huber, R.~Alkofer and S.~P. Sorella,  arXiv:0910.5604 [hep-th] (2009), to app. in \emph{Phys. Rev.} \textbf{D}.

\bibitem[Kugo and Ojima(1979)]{Kugo:1979gm} 
T.~Kugo and I.~Ojima, \emph{Prog. Theor. Phys. Suppl.} \textbf{66}, 1 (1979).

\bibitem[Kugo(1995)]{Kugo:1995km} 
T.~Kugo,  arXiv:hep-th/9511033 (1995).

\bibitem[Dudal et~al.(2009{\natexlab{a}})]{Dudal:2009xh} 
D.~Dudal, S.~P. Sorella, N.~Vandersickel and H.~Verschelde, \emph{Phys. Rev.} \textbf{D79}, 121701 (2009{\natexlab{a}}).

\bibitem[Cucchieri and Mendes(2007)]{Cucchieri:2007md}
A.~Cucchieri and T.~Mendes, \emph{PoS} \textbf{LAT2007}, 297 (2007).

\bibitem[Dudal et~al.(2008)]{Dudal:2008sp}
D.~Dudal, J.~A. Gracey, S.~P. Sorella, N.~Vandersickel and H.~Verschelde,  \emph{Phys. Rev.} \textbf{D78}, 065047 (2008).

\bibitem[Maggiore and Schaden(1994)]{Maggiore:1993wq} 
N.~Maggiore and M.~Schaden, \emph{Phys. Rev.} \textbf{D50}, 6616 (1994).

\bibitem[Dudal et~al.(2005)]{Dudal:2005na}
D.~Dudal, R.~F. Sobreiro, S.~P. Sorella and H.~Verschelde, \emph{Phys. Rev.}  \textbf{D72}, 014016 (2005).

\bibitem[Sorella(2009)]{Sorella:2009vt}
S.~P. Sorella, \emph{Phys. Rev.} \textbf{D80}, 025013 (2009).

\bibitem[Baulieu et~al.(2009)]{Baulieu:2009ha}
L.~Baulieu, D.~Dudal, M.~S. Guimaraes, M.~Q. Huber, S.~P. Sorella,  N.~Vandersickel and D.~Zwanziger,  arXiv:0912.5153 [hep-th]  (2009).

\bibitem[Dudal et~al.(2009{\natexlab{b}})]{Dudal:2009zh}
D.~Dudal, S.~P. Sorella, N.~Vandersickel, and H.~Verschelde, \emph{JHEP} \textbf{08}, 110 (2009{\natexlab{b}}).



\end{thebibliography}


\end{document}